\documentstyle[psfig]{article}
\begin{document}

\begin{center}
\LARGE \bf 
The High-Redshift Supernova Search -- Evidence for a Positive 
Cosmological Constant\footnote{To appear in Dark'98, eds. H.
Klapdor-Kleingrothaus and L. Baudis, Singapore: World Scientific
Publishing}

\vspace*{10mm}
\large

Bruno Leibundgut, Gertrud Contardo, Patrick Woudt and Jason
Spyromilio\footnote{We report results from the High-z Supernova Search Team.
The team includes  B.P.~Schmidt (MSSSO), M.~Phillips (LCO), 
N.~Suntzeff, R.~Schommer and C. Smith (CTIO), A.~Clocchiatti (Universidad 
Catolica, Santiago), M.~Hamuy (Steward Obs.), 
R.~Kirshner, P.~Garnavich, S. Jha and P.~Challis (Harvard Univ.), 
C.~Hogan, C.~Stubbs,
A.~Diercks and D.~Reiss (Univ. Washington), A.~Filippenko and A.~Riess 
(Univ. California, Berkeley), R.~Gilliland (STScI), J.~Tonry (Hawaii).
More information is available at {\tt
http://cfa-www.harvard.edu/cfa/oir/Research/supernova/HighZ.html}}
\vspace*{2mm}

European Southern Observatory, Karl-Schwarzschild-Strasse 2,
D-85748 Garching, Germany
\vspace*{5mm}

\normalsize
\section*{Abstract}

\begin{minipage}{120mm}
A new component of the Universe which leads to an accelerated cosmic expansion
is found from the measurements of distances to high-redshift type Ia
supernovae. We describe the method and the results obtained from the
observations of distant supernovae. The dependence on the understanding
of the local type Ia supernovae is stressed. The lack of a good
understanding of the stellar evolution leading to the explosion of the 
white dwarf,
the exact explosion physics and the current difficulties in calculating
the emission from the ejecta limit the theoretical support.
Despite the
current ignorance of some of the basic physics of the explosions, the
cosmological result is robust. The empirical relations seem to hold for
the distant supernovae the same way as for the local ones and the
spectral appearance is identical. The distances to the high-redshift
supernovae are larger than expected in a freely coasting, i.e. empty,
Universe. A positive cosmological constant is inferred from these
measurements.
\end{minipage}

\end{center}

\section{Introduction}

The quest to measure the global dynamics of the Universe has dominated
cosmological observations since the discovery of cosmic expansion.
Observational tests have been devised ever since, but only recently have
the measurements achieved an accuracy which allows us to draw more 
definite conclusions. Evidence has been accumulated from
the observations of the mass concentrations on the largest scales and
the time required to build them up from the earliest imprints in the
cosmic microwave background (Bahcall \& Fan 1998), from the
fact that the formation of old stellar systems are incompatible with the
Hubble time as derived from the present-day expansion rate, the Hubble
constant (e.g. Carroll et al. 1992, Sandage 1988), and
finally from the first direct distance measurements at redshifts larger
than 0.1. 

It is through the use of Type Ia supernovae that we now are able to glimpse
at what the global energy density of the Universe may be and what is
governing the expansion field. The intriguing new results are quite
contrary to the expectations and will need a lot more scrutiny.
Ultimately, the result will have to be supported by independent
measurements of the changes of the expansion rate of the Universe. 

If the distances derived from the supernovae are indeed larger than
expected in a freely coasting Universe, i.e. $q_0$=0, then a new
component to the cosmic energy budget has to be added (Schmidt et al.
1998, Garnavich et al.
1998a, Riess et al. 1998, Perlmutter et al. 1998, 1999). This is commonly
expressed as a cosmological constant, but could be in a more general
form (White 1998, Garnavich et al. 1998b). This new form of energy is 
a different, independent component to the dark matter, which was the 
topic of this conference.

We will describe how type Ia supernovae (SNe~Ia) can be used to determine
cosmological distances and point out the caveats with the current
measurements (\S 2). The High-z Supernova Search Team has
presented results of the first two years observing with astonishing
conclusions. The Berkeley Supernova Cosmology Project independently 
arrived at the
same conclusions through the same technique, but a completely
independent data sample. Their results are presented in this volume
by Isobel Hook and Pilar Ruiz-Lapuente. We will concentrate on the results
found by the High-z Team in section 3. Potential pitfalls of the
measurement are presented in section 4. We finish with a brief discussion of 
the implications of an accelerated expansion and conclusions (\S 5). 

\section{Type Ia supernovae in cosmology}

\subsection{Probing the nature of cosmological redshifts through time dilation}

The regular temporal behavior of SNe~Ia provides a simple, yet
important, test of the basic interpretation of redshift as due to cosmic
expansion. The most stringent indication of this has been the uniformity of
the cosmic microwave background (Mather et al. 1990, Peebles et al.  1991).
A direct proof for the expansion, however, had been missing but is provided
by a clock at high redshift. The SN~Ia light curves have been proposed as
such a clock and can also be used to search for any evolutionary effects
in the explosion (Wilson 1939,
Colgate 1979, Tammann 1979, Leibundgut 1990). The fundamental nature of
the cosmological redshift can be probed with a single distant supernova
and the assumption that it does not differ significantly from nearby
ones. Such an analysis became possible for the first time with light
curves of SN~1995K at
a redshift of 0.48 (Leibundgut et al. 1996) and a small sample of five 
supernovae from the Supernova Cosmology Project (Goldhaber et al. 1997).
The same effect has been observed in the spectral evolution of the distant
SN~1996bj (Riess et al. 1997).
The result strikingly demonstrates the conventional interpretation of
redshift being an effect of the cosmic expansion
rather than any theories involving connections of redshift with an
energy loss of the photon. SN~1995K could not be explained in a 
non-expanding Universe unless it would have had unprecedented attributes 
(Leibundgut et al. 1996). In particular, the light curve shape would
have made it the slowest declining SN~Ia observed ever with a spectrum basically
indistinguishable from local SNe~Ia (Schmidt et al. 1998). 
This is contrary to all correlations found in the local sample.
Nonetheless, other interpretations of these observations have been advanced 
as well (Narlikar \& Arp 1997, Segal 1997). 

\subsection{Cosmological distances from standard candles}

Distance measurements from SNe~Ia are made through a modified standard
candle scheme, where luminosity distances are derived. This is a very
simple test of the global geometry which has been proposed for several 
decades (Heckmann
1942, Robertson 1955, Hoyle \& Sandage 1956, Sandage 1961) for a number
of standard candle candidates. The assumption in this method is that the
luminosity evolution of the standard candle is negligible or at least can be
measured accurately. For SNe~Ia it is generally assumed that their
maximum luminosity does not change as a function of cosmic age. We will discuss
this assumption below (\S~\ref{evol}). For an exact standard candle the
cosmological parameters are described in the implicit equation

\[D_L = \frac{(1+z)c}{H_0|\kappa|^{1/2}} \;\; S\!\left\{
\begin{array}{c} \rm 
\\ \rm \end{array} \right.\!\!\!\!\!\!\!|\kappa|^{1/2} \int_{0}^{z}
[\kappa(1+z^{\prime})^2 + \Omega_M(1+z^{\prime})^3 +
\Omega_{\Lambda}]^{-1/2} dz^{\prime}\left\} \begin{array}{c} \rm \\ \rm
\end{array} \right.\!\!\!\!\!\!\! \]

\noindent (e.g. Carroll et al. 1992). Here $\Omega_M =
\frac{8\pi G}{3H_0^2}\rho_M$ stands for the matter content, which
depends 
only on the mean matter density of the universe
$\rho_M$, and $\Omega_{\Lambda} =
\frac{\Lambda c^2}{3H_0^2}$ describes the
contribution of a cosmological constant to the expansion factor. 
$\kappa$ is the curvature term and obeys 
\[\kappa = 1 - \Omega_M - \Omega_{\Lambda}. \]
The integration provides the cosmological distance element out to the
source redshift $z$. 
 
$S(\chi)$ takes the form
  
\[S(\chi) = \left\{ \begin{array}{lll}
                    \sin(\chi) & & \kappa < 0 \\
                    \chi & {\rm for} & \kappa = 0 \\
                    \sinh(\chi) & & \kappa > 0.
                    \end{array}
            \right. \]
                                                            
The change of the expansion rate, usually denoted as the deceleration
parameter $q_0$, is defined as $q_0 = \frac{\Omega_M}{2} - \Omega_{\Lambda}$.

The supernova distances are measured as the distance modulus

\[m-M=5\log(D_L)+25 \] 

\noindent with the luminosity distance in units of megaparsecs. The
most probable values of the cosmological parameters are then found in a
least squares fit, possibly assuming certain boundary conditions 
(Riess et al. 1998, Perlmutter et al. 1998, Leibundgut 1998). 
It has to be noted that the
present-day value of the Hubble constant, $H_0$, is not of relevance in the 
determination of
the energy density of the Universe, but rather depends on the zero-point
which is derived from the nearby supernovae. The deceleration is
entirely measured from the apparent magnitude differences between the
nearby sample and the distant supernovae. With a sufficiently large
redshift range the degeneracy between $\Omega_M$ and $\Omega_{\Lambda}$
can be broken by standard candles (Goobar \& Perlmutter 1995). A much
more effective way, however, is to find a measurement which depends on the
cosmological parameters in a different way. This can be achieved by the
comparison of the supernova result with measurements of the cosmic
microwave background (White 1998, Eisenstein et al. 1998, Garnavich et al.
1998b).

\subsection{Type Ia Supernovae as standard candles}

Recent years have seen a dramatic increase in observational material on
SNe~Ia. Well-sampled light curves in many filters have been assembled 
for about 50 nearby
supernovae ($z<0.1$; Hamuy et al. 1996a, Riess et al. 1999). The first
secure absolute distances from direct Cepheid measurements of SNe~Ia
have confirmed that they exhibit a very small scatter in their maximum
light luminosity (Saha et al. 1998, Tammann in this volume). These
results are used to refine our understanding of the explosive events. 
The most important result for cosmological applications of SNe~Ia is the 
nearly uniform luminosity and the possibility to correct for variation
in the peak luminosity by a distance independent parameter, i.e. the
decline from maximum during the first two weeks (Phillips et al. 1993,
Hamuy et al. 1996b, Riess et al. 1996, 1998). It seems that SNe~Ia
can be described fairly well as a one parameter family as many different
parameters correlate with the decline rate. The decline rate, usually
denoted as $\Delta m_{15}$ for the $B$ band, correlates with the peak
luminosity (Hamuy et al. 1996a, Riess et al. 1996), the expansion
velocity of the ejecta (Mazzali et al. 1998), the color at maximum light
(Hamuy et al. 1996b, Riess et al. 1996, Branch 1998), the galaxy type
(Hamuy et al. 1996b, Riess et al. 1996, 1999), line ratios of certain
elements (Nugent et al. 1995), and possibly with the late-decline rate
of individual filter light curves (Hamuy et al. 1996c). Most of these
correlations have been established for $B$ and $V$ filters. The
correction to the luminosity at maximum is at heart of the use of SNe~Ia
to measure cosmological distances.

\begin{figure}[h]
\psfig{file=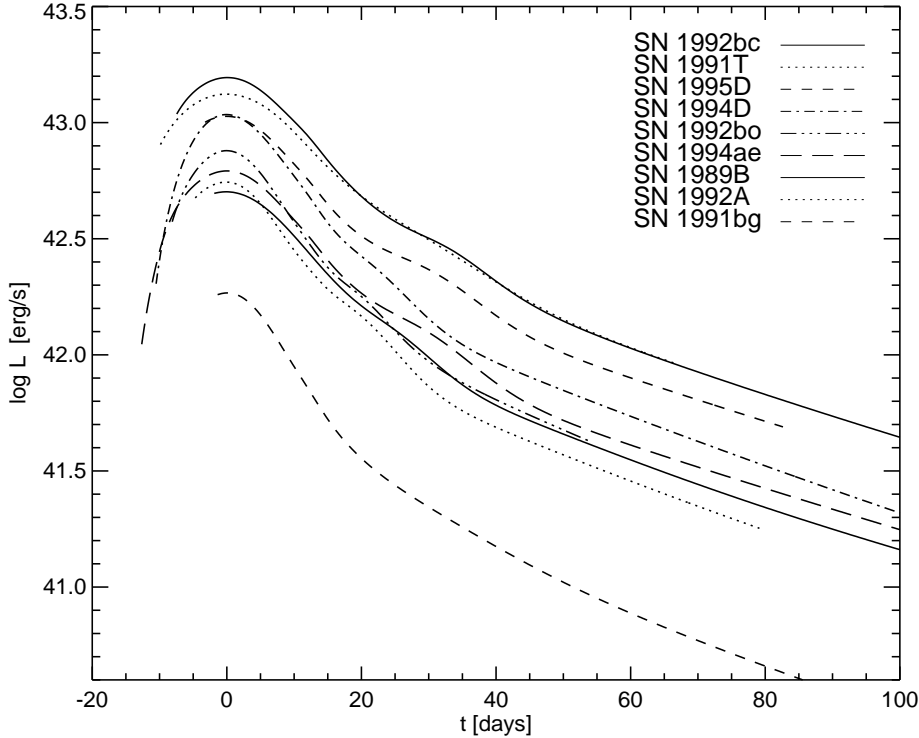,angle=0,width=140mm,%
bbllx=0mm,bblly=125mm,bburx=200mm,bbury=270mm}
\caption{Bolometric light curves of nearby SNe~Ia (Contardo et al.  1999).}
\label{fig:bol}
\end{figure}

Despite these well-established correlations some questions remain as to
the exact nature of these explosions. The bolometric light curves constructed
from the optical data do not show such a nice correlation with the
decline rate. Fig.~\ref{fig:bol} displays a set of bolometric light
curves of well-observed, nearby SNe~Ia. It is obvious that the second maximum
observed in the $I$ (Suntzeff 1996, Ford et al. 1993) and the
near-infrared light curves (Elias et al. 1985) also
appears in the bolometric light curves (Contardo et al. 1999). This
was originally pointed out by Suntzeff (1996) for the bolometric light
curve of SN~1992A. The strength of this inflection
varies between individual events. A clear trend of luminosity and
strength of the inflection is not detected and there seems to be another
parameter governing the energy release from the fireball. Surprisingly
the decline of the bolometric light curve of SN~1991bg does not differ
from the one of the other SNe~Ia. This is in marked contrast to the
filter light curves of this supernova. The $B$ and $V$ light curves of
SN~1991bg declined much faster than for any other known SN~Ia (Leibundgut et al.
1993, Turatto et al. 1996). 

It should also be stressed that we currently do not understand the
physics for the decline -- luminosity relation (see, however, H\"oflich
et al. 1996). The hydrodynamics and
the radiation transport of the SN~Ia ejecta is fairly uncertain and a
number of models have been proposed (Arnett \& Livne 1994, 
Woosley \& Weaver 1994, Khokhlov et al. 1993, H\" oflich \& Khokhlov 1996).
Supernova atmospheres are far
from thermal equilibrium as demonstrated by the lack of emission in
spectral regions with none or few emission lines (Spyromilio et al.
1992) or the occurrence of maximum luminosity in different filter bands
(Contardo et al. 1999). 

The exact stellar evolution which leads to the progenitors of SNe~Ia is
not understood and a variety of astronomical objects has been proposed
(cf. Branch et al. 1995). Also the explosion physics have not been
solved yet. A number of explosion models has been proposed, but
observational distinctions have eluded us so far. These uncertainties
have to be addressed in any serious application of SNe~Ia for cosmology.

\section{Evidence from distant supernovae for a cosmological constant}

The High-z Supernova Team was formed in 1994 to pursue the observations
of distant, i.e. $z>0.3$, supernovae. The team members have access to
almost every major telescope and are located on four different continents.
The current tally is at 116 candidate supernovae discovered with 45
spectroscopically confirmed SNe~Ia (Woudt et al. 1999). 
We will concentrate
on the first ten fully reduced objects as published by Riess et al.
(1998). 

A number of corrections have to be applied to the data before
luminosity distances can be derived. Technical problems include the
accurate photometry as most supernovae are very faint. 
The photometry then has to be converted to
the rest frame of the supernova. This K-correction is not only a function
of redshift and phase, but also depends on the decline rate and spectral
appearance of the supernova. 
All light curves are then corrected for time dilation by dividing the
phases with $(1+z)$.
The reddening of the distant supernovae is
determined from the intrinsic rest frame color. A reddening correction is
applied either implicitly in the multi-color light curve shape method 
(Riess et al. 1996) or as an extra step when the $\Delta m_{15}$
procedure is used. 
Indeed, only one or two objects in our sample show significant 
reddening (Riess et al. 1998). Finally, a correction for the light curve
shape is applied to the distant supernovae. The correlation establishing
this correction is based on a large sample of nearby supernovae (Hamuy
et al. 1996a, Riess et al. 1999). It is important to realize that the
relation has to be derived for the local sample in exactly the same way as
it is applied to the distant set. In particular, the same filter set and
also the same phase range have to be applied, in order not to introduce
additional parameters which are not measured for the distant supernovae
(Riess et al. 1998). Interestingly, the significance of the derived result 
strongly depends on the control of the local sample (Riess et al. 1998,
Leibundgut 1998). A detailed discussion of these corrections and their
accuracies is given in section~\ref{tech}.

\begin{figure}[t]
\psfig{file=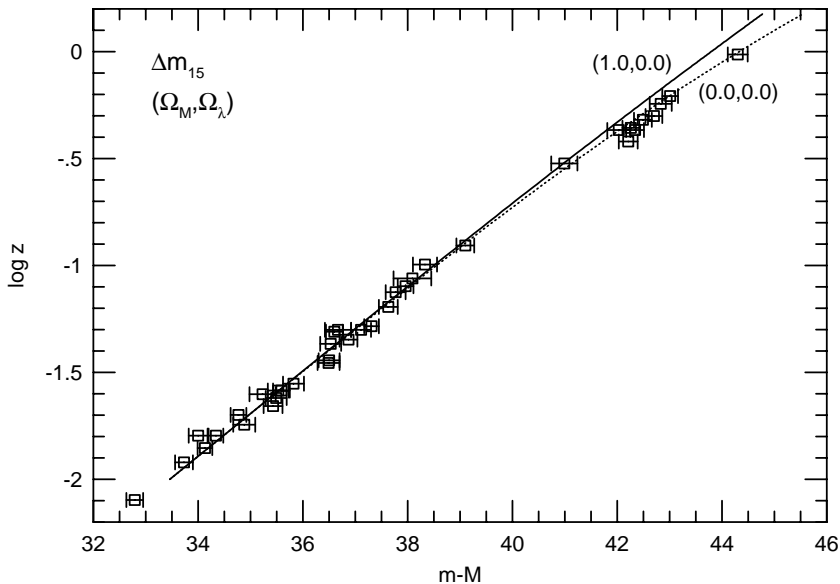,angle=0,width=140mm,%
bbllx=-40mm,bblly=25mm,bburx=270mm,bbury=200mm}
\caption{Hubble diagram of local and distant supernovae (data from Riess
et al. 1998). The curves indicate the distance vs. redshift evolution in
an Einstein-de~Sitter (full line) or an empty Universe (dotted line).}
\label{fig:hub}
\end{figure}

For each supernova a luminosity distance is derived in this way.
Comparison to theoretical models is then made through a modified Hubble
diagram where the luminosity distance, i.e. distance modulus, 
is plotted vs. the redshift (Fig.~\ref{fig:hub}). The local supernovae
determine the locus of all models in the linear expansion regime (out to
$z\approx 0.1$). It is the relative distances of the high-z supernovae
compared to the local sample which provides the information on the
change in the expansion. 

\begin{figure}[h]
\psfig{file=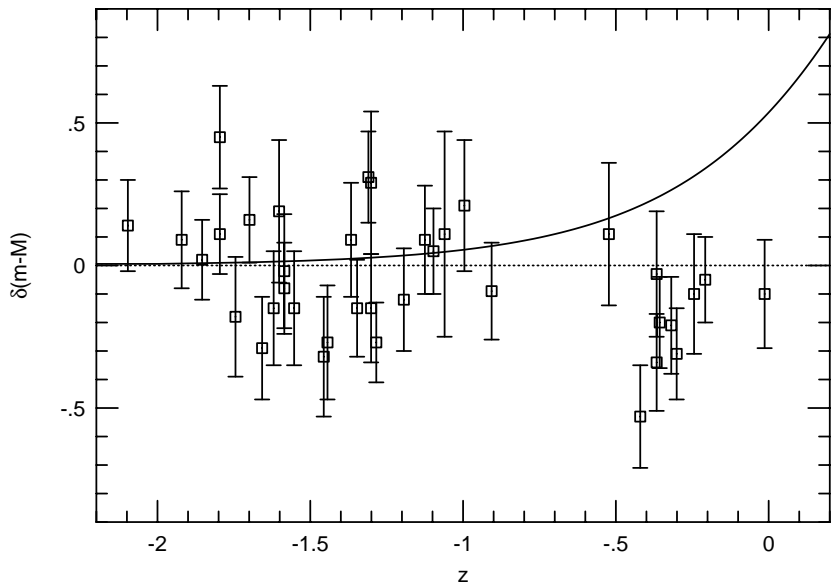,angle=0,width=140mm,%
bbllx=-40mm,bblly=25mm,bburx=270mm,bbury=200mm}
\caption{Distance differences from the individual supernovae from the
expected line in an empty Universe.}
\label{fig:rel}
\end{figure}

It is evident in this figure that the data for the distant SNe~Ia
do not follow either model plotted. They are compared to the
Einstein-de~Sitter models ($\Omega_M=1, \Omega_{\Lambda}=0$) and an
empty Universe without matter and no contribution of a cosmological
constant ($\Omega_M=0, \Omega_{\Lambda}=0$). While the
Einstein-de~Sitter model is clearly ruled out, a model with no mass
density and no cosmological constant is also not a good fit. The distant
supernovae all lie below, i.e. at larger distances, than the expectations
from these models. This becomes even clearer when the Hubble diagram is
normalized to the empty Universe model (Fig.~\ref{fig:rel}). 

The distant SNe~Ia show a systematic trend towards distances which appear
to be even larger than in a freely coasting universe. 
The distant SNe~Ia are about 0.2 magnitudes from the dividing
line of an empty Universe. If the
luminosity of SNe~Ia has not changed since $z\approx 0.5$, the
mean of the redshift distribution, then we are forced to admit that the
distances of these objects have to be larger than in a non-decelerated
Universe, and an acceleration has boosted the distances. The most
obvious candidate for such an acceleration is a positive cosmological 
constant. 

Of course, the caveat of any luminosity evolution
or other subtle systematic effects have to be critically reviewed before
this result can be accepted. We will come back to these issues in
section~\ref{evol}.

The luminosity distances are most sensitive to the combination $\Omega_M
- \Omega_{\Lambda}$ (e.g. White 1998, Eisenstein et al. 1998). Thus, the
supernovae determine an uncertainty region nearly perpendicular to the flat
geometry solutions. For our data set we find that any solutions for a 
positive matter
density require also a contribution of a cosmological constant. The
confidence intervals are given in Table~\ref{tab:riess}. We do not
have a large enough redshift range for an accurate determination of
$\Omega_M$ and $\Omega_{\Lambda}$ independently.
Despite this current limitation we find a very
high confidence limit for a cosmological constant and even an
acceleration of the universal expansion since the time the SNe~Ia
exploded. 

\begin{table}[t]
\begin{center}
\footnotesize\rm
\caption{Summary of the results on the cosmological parameters from SNe~Ia}
\label{tab:riess}
\begin{tabular}{l c c c c c c c}
\hline
\hline
Method & $\Omega_M$ & $\Omega_{\Lambda}$ & age &
P($\Omega_{\Lambda}>0$) & P($q_0 < 0$) \\
 & & & ($10^9$ years) & & \\
\hline
MLCS & $\ldots$ & $\ldots$ & 14 & 2.9$\sigma$ & 2.4$\sigma$ \\
$\Delta M_{15}$ & $\ldots$ & $\ldots$ & 15 & 3.9$\sigma$ &
3.8$\sigma$ \\
MLCS+SN 1997ck & 0.00$^{+0.60}_{-0.00}$ & 0.48$^{+0.72}_{-0.24}$ &
14 & 2.8$\sigma$ & 2.7$\sigma$ \\
$\Delta M_{15}$+SN 1997ck & 0.72$^{+0.44}_{-0.56}$ & 1.48$^{+0.56}_{-0.68}$ &
15 & 3.8$\sigma$ & 3.7$\sigma$ \\
\hline
\hline
\end{tabular}
\end{center}
\end{table}

\section{Systematic error sources}

There are number of effects which possibly could alter the result we
found in the previous section. They can be roughly divided into two
classes. Technical impediments, like accurate photometry, K-corrections,
light curve sampling, sample
contamination, and selection effects can be controlled by adequate
observing and reductions strategies. Fundamental problems arise from
possible gravitational amplification or de-amplification, absorption,
and evolution of SNe~Ia explosions. 

\subsection{Corrections to the observed data}
\label{tech}

Accurate photometry is a pre-requisite for the determination of the peak
brightness of the objects. The most important contaminant is the
background light from the galaxy on which the supernova is superposed.
In many cases,
the underlying host galaxy light has to be carefully subtracted. As the
supernova fades these problems are exacerbated. This is particularly
important as the decline rate is determined by the later phase, i.e.
fainter, observations. 
The individual photometry points are then corrected for the effects of
the redshift. The observed flux has to be converted to a rest frame
magnitude to be comparable to the local SNe~Ia. These K-corrections are
time-dependent and have to be determined from nearby supernovae over the
whole range of the light curve. A slight dependence of the K-corrections
on the decline rate, i.e. the intrinsic color, of the supernova has been
noted (Riess et al. 1998) and has to be included. Despite the
complications in this process a very good accuracy has been obtained
(e.g. Schmidt et al. 1998), as the problem can be controlled very well
and only depends on the availability of sufficient data on nearby
supernovae. 

Since the peak magnitudes are required to determine the decline
parameters and the distances, a well-sampled light curve is needed for each
supernova. Global observing campaigns are organized to achieve this
goal. The critical observations near maximum and about two weeks after
maximum, which are the most crucial for the determination of the peak
magnitude and the decline-rate corrections, are virtually guaranteed by
the search technique for the distant supernovae 
(Perlmutter et al. 1997, Schmidt et al. 1998). The sample of local
supernovae is large enough that we can select suitable subsamples (Hamuy
et al. 1996b, Riess et al. 1999). For well-sampled light curves the
errors of the light curve parameters are significantly reduced.

A critical evaluation of each object in the sample for its supernova
type is unavoidable. Since the classification for supernovae is based
on spectroscopy near maximum light (Harkness \& Wheeler 1990, Filippenko
1997), it is imperative to obtain a
spectrum of each object. Occasionally the spectrum is not decisive enough
and even the combination with the light curve can not exclude
ambiguities. In the sample discussed here we have one object for which
it was not possible to unambiguously determine its classification (Riess
et al. 1998). For another object, SN~1997ck at $z=0.97$, it was not possible
to obtain a spectrum. Exclusion of these objects from the sample does
not change our results (Table~\ref{tab:riess}). 

Selection effects could change the result. A recent discussion of the
Malmquist bias by Teerikorpi (1998) demonstrates that such an effect, if
not detected, could yield to an underestimate of the deceleration. 
In effect, since
there is some intrinsic scatter for every standard candle, the volume
which is sampled for a standard candle at a given redshift is biased 
to a larger distance than
indicated by the straight mean magnitude and the distances are
over-estimated. Basically, the volume sampled at $m+\sigma m$ is
larger than the one with $m-\sigma m$, where $\sigma m$ is the scatter
of the standard candle. This effect works in all magnitude limited samples. For
the distant supernovae one also has to consider that some objects may have
been missed by the search. Fortunately, the scatter of the nearby sample of
SNe~Ia shows such a small range (0.15 mag, Schmidt et al. 1998)
that the expected systematic errors are still smaller than the total
offset measured for the distant SNe~Ia.

\subsection{Astrophysical influences}
\label{evol}

Gravitational lensing of distant objects is unavoidable. Most
importantly the apparent brightness can be changed due this effect.
Since gravitational amplification is wavelength independent it can not be
detected in the objects' light directly. For SNe~Ia this is the only
effect which can not be inferred from the SN observations alone.
Mapping of the gravitational potential along the line of sight is
required (e.g. Wambsganss et al. 1998). The redshift out to which
SNe~Ia have been observed so far is not large enough to suffer from any
significant influence from gravitational lensing (Wambsganss et al.
1997, Holz 1998). Even in the most extreme case where all matter is clumped
('empty beam') our result of an accelerated
expansion will not change significantly (Holz 1998). 

A possible explanation of the apparent faintness of the distant
supernovae could be absorption. All observations are corrected for
absorption by the Galaxy. Observing in two filter bands should allow us 
to detect absorption in the host galaxy. This implicitly assumes that
the intrinsic color evolution of all SNe~Ia can be traced and that the
reddening law is the same at $z=0.5$ as in our Galaxy. Any absorption at
other redshifts is not considered. The average column density as
measured from QSO absorbers out to $z\approx 0.5$ is small and can be
ignored. Most distant SNe~Ia have a very small absorption. This is due
to two selection effects. First, heavily absorbed supernovae are less
likely to be discovered and, second, the spectroscopic follow-up
observations concentrate mostly on SN candidates well separated from the
galaxies to avoid strong contamination from the galaxy light. Thus, we
do expect rather small reddening for most of the distant supernovae.
It is unlikely that dust extinction systematically affects the distances
to mimic the observations. This is because, significant absorption
along certain sight lines would increase the scatter in the observed
distances. This is not observed (Riess et al. 1998). 

A most critical assumption is the equivalency of the distant supernovae
to the local ones. Any evolution of the supernovae as a function of,
e.g. progenitor age, could influence the peak luminosity of the light
curve. After all, the distant supernovae exploded some 5~Gyr earlier
than the ones in the local Universe. The lack of detailed explosion
models currently prevents a robust theoretical prediction. An attempt
was made to investigate the influences of a number of parameters on the
light curves (H\"oflich et al. 1998). The strongest effect found was the
chemical composition of the progenitor star which could change the blue
part of the optical spectrum. The influence is rather small, however. An
empirical test for the similarity of the distant SNe~Ia with the local
ones is by comparing the observational properties of the distant sample
to the nearby one. The currently available spectroscopy has not detected
any significant deviations (Riess et al. 1998, Perlmutter et al. 1998).
In most cases, the spectra are not of high enough quality to guarantee
this result, but the few objects with excellent spectroscopy show the
same evolution as for nearby SNe~Ia (Riess et al. 1997, 1998,
Perlmutter et al. 1998). The rest-frame colors and the light curve shapes
of the distant objects also do not deviate. We are faced with the
possibility that all measurable distance independent quantities of SNe~Ia
appear unchanged, while the luminosity could have changed. This is a
rather unlikely proposition, but will need more critical scrutiny. 

Possible sample differences, as typically produced by a selection bias,
have to investigated as well. The global sample properties of the local
and distant sample have been found to be very similar (Riess et al.
1998, Leibundgut et al. 1999). The good standard candle quality of
SNe~Ia is a very important asset in this respect. 

\section{Discussion and Conclusions}

Distant SNe~Ia provide striking evidence for an acceleration of the
universal expansion over the last $\sim 6$~Gyr. Commonly such an
acceleration has been identified by the possible contribution of a
vacuum density, i.e. cosmological constant. This is the first clear
indication for the existence of a significant vacuum density. The
cosmological constant acts like a negative pressure term for the
expansion. It is possible to examine the equation of state of the
Universe in more detail by splitting the contributions to the geometric
term in the equation for the luminosity distance into the several components 
(Garnavich et al. 1998b). By doing so the dominant source of the
acceleration can be determined. The best fit to the data is
using a component very similar to the cosmological constant where
pressure is proportional to the negative density ($P\propto -\rho$).
Topological defects could also be responsible. A network of
non-commuting cosmic strings would have an average effective relation of
$P\propto -\frac{\rho}{3}$, but are excluded for any flat spatial
geometry ($\Omega_{\rm total}=1$). 

It is interesting that the best cosmological parameters found in our
study provide an age estimate of the Universe which is in agreement
with the oldest stellar components. The age we find
(Table~\ref{tab:riess}) is about 14~Gyr, which comfortably includes the
globular cluster ages (Riess et al. 1998). The fact that the Universe
has suffered from an accelerated expansion means that the simple extrapolation
based on the present-day value of the Hubble constant underestimates the
age of the Universe.

As mentioned in the introduction, the combination of the supernova
result with accurate measurements of the cosmic microwave background (CMB)
fluctuations provides the possibility to restrict the allowed range in
the $\Omega_M - \Omega_{\Lambda}$ plane considerably (White 1998,
Eisenstein et al. 1998). The two measurements are almost orthogonal 
in this parameter space. A first attempt was made by White
(1998) and Garnavich et al. (1998b) combining the supernova data with
the constraints found on the CMB (Hancock et al. 1997). The combined
constraints give a narrow region around $\Omega_M=0.3$ and
$\Omega_{\Lambda}=0.7$ with 3$\sigma$ confidence limits reaching from about
$-0.2<\Omega_{\Lambda}<1.3$ and $\Omega_M < 1.0$ (Garnavich et al.
1998b). The Einstein-De~Sitter model is excluded by many $\sigma$. 

A number of distant SNe~Ia have been already observed and are about to
be analyzed and published. The Berkeley Supernova Cosmology project is
publishing a large set of supernovae (Hook, this volume, Perlmutter et
al. 1999). The High-z team expects to have about 20 additional SNe~Ia
analyzed next year. The supernova sample has increased to a size where
statistical uncertainties are not important any more. It is the
systematic error sources described in section~\ref{tech} and \ref{evol}
which dominate the uncertainty in the measurements. They will have to be
tackled one by one. The most important seems currently the question of
evolution. We need a much better understanding of the spectral evolution
of SNe~Ia at high redshift. This implies that spectroscopy not just for
the classification of the object, but for a detailed spectral
analysis will have to be obtained. In addition, programs to investigate 
the environment of the distant
supernovae and compare them to the local sample are under way. They
include the study of the parent galaxy morphology and metalicities. 

The CMB constraints will improve with the future space missions of MAP
and PLANCK. There are excellent prospects that the value of $\Omega_M$ 
and $\Omega_{\Lambda}$ will be
pinned down fairly accurately in less than 10 years from now. 

\section*{References}
\begin{list}{}%
{\setlength {\itemindent -10mm} \setlength {\itemsep 0mm} \setlength 
{\parsep 0mm} \setlength {\topsep 0mm}}
\item Arnett, W. D. \& Livne, E. 1994, ApJ, 427, 315
\item Bahcall, N. A. \& Fan, X. 1998, ApJ, 504, 1
\item Branch, D. 1998, ARA\&A, 36, 17
\item Branch, D., Livio, M., Yungelson, L. R., Boffi, F. R., \& Baron, E. 1995, PASP, 107, 1019
\item Carroll, S. M., Press, W. H., \& Turner, E. L. 1992, ARA\&A, 30, 499
\item Colgate, S. A. 1979, ApJ, 232, 404
\item Contardo, G., Leibundgut, B., \& Vacca, W. D. 1999, in preparation
\item Eisenstein, D. J., Hu, W., \& Tegmark, M. 1998, ApJ, 504, L57
\item Elias, J. H., Matthews, K., Neugebauer, G., \& Persson, S. E. 1985, ApJ, 196, 379
\item Filippenko, A. V. 1997, ARA\&A, 35, 309
\item Ford, C. H., Herbst, W., Richmond, M. W., Baker, M. L., Filippenko, A. V., Treffers, R. R., Paik, Y., \& Benson, P. J. 1993, AJ, 106, 1101
\item Garnavich, P. M., et al. 1998a, ApJ, 493, L53
\item Garnavich, P. M., et al. 1998b, ApJ, 509, in press (astro-ph/9806396)
\item Goldhaber, G., et al. 1997, Thermonuclear Supernovae, eds. P. Ruiz-Lapuente, R. Canal, \& J. Isern, Dordrecht: Kluwer, 777
\item Goobar, A \& Perlmutter, S. 1995, ApJ, 450, 14
\item Hamuy, M., et al. 1996a, AJ, 112, 2408
\item Hamuy, M., et al. 1996b, AJ, 112, 2398
\item Hamuy, M., Phillips, M. M., Suntzeff, N. B., Schommer, R. A., Maza, J., Smith, R. C., Lira, P., \& Avil\' es, R. 1996c, AJ, 112, 2438
\item Hancock, S., Rocha, G., Lasenby, A. N., \& Guti\'errez, C. M. 1998, MNRAS, 294, L1
\item Harkness, R. P., \& Wheeler, J. C. 1990, Supernovae, ed. A. G. Petschek, (New York: Springer), 1
\item Heckmann, O. 1942, Theorien der Kosmologie, Berlin: Springer
\item H\"oflich, P. \& Khokhlov, A. 1996, ApJ, 457, 500
\item H\"oflich, P., Khokhlov, A., Wheeler, J. C., Phillips, M. M., Suntzeff, N. B., \& Hamuy, M. 1996, ApJ, 472, L81
\item H\"oflich, P., Wheeler, J. C., Wheeler, \& Thielemann, F.-K. 1998, ApJ, 495, 617
\item Holz, D. E. 1998, ApJ, 506, L1
\item Hook, I., this volume
\item Hoyle, F. \& Sandage, A. 1956, PASP, 68, 30
\item Khokhlov, A., M\"uller, E., \& H\"oflich, P. 1993, A\&A, 270, 223
\item Leibundgut, B. 1990, A\&A, 229, 1
\item Leibundgut, B. 1998, Supernovae and Cosmology, eds. L. Labhardt, B. Binggeli, R. Buser, Basel: University of Basel, 61
\item Leibundgut, B., et al. 1993, AJ, 105, 301
\item Leibundgut, B., et al. 1996, ApJ, 466, L21
\item Leibundgut, B., Schmidt, B. P., Spyromilio, J., \& Phillips, M. M. 1999, Looking Deep in the Southern Sky, eds. R. Morganti and W. Couch, in press
\item Mather, J., et al. 1990, ApJ, 354, L37
\item Mazzali, P. A., Cappellaro, E., Danziger, I. J., Turatto, M., Benetti, S. 1998, ApJ, 499, L49
\item Narlikar, J. V. \& Arp, H. C. 1997, ApJ, 482, L119
\item Nugent, P., Phillips, M. M., Baron, E., Branch, D., \& Hauschildt, P. 1995, ApJ, 455, L147
\item Peebles, P. J. E., Schramm, D. N., Turner, E. L., \& Kron, R. G. 1991, Nature, 352, 769
\item Perlmutter, S., et al. 1997, ApJ, 483, 565
\item Perlmutter, S., et al. 1998, Nature, 391, 51
\item Perlmutter, S., et al. 1999, ApJ, submitted
\item Phillips, M. M. 1993, ApJ, 413, L105
\item Riess, A. G., Press, W. M., \& Kirshner, R. P. 1996, ApJ, 473, 88
\item Riess, A. G., et al. 1997, AJ, 114, 722
\item Riess, A. G., et al. 1998, AJ, 116, 1009
\item Riess, A. G., et al. 1999, AJ, in press (astro-ph//9810291)
\item Robertson, H. P. 1955, PASP, 67, 82
\item Ruiz-Lapuente, P., this volume
\item Saha, A., Sandage, A., Labhardt, L., Tammann, G. A., Macchetto, F. D., \& Panagia, N. 1997, ApJ, 486, 1
\item Sandage, A. 1961, ApJ, 133, 355
\item Sandage, A. 1988, ARA\&A, 26, 561
\item Schmidt, B. P., et al. 1998, ApJ, 507, 46
\item Segal, I. E. 1997, ApJ, 482, L115
\item Spyromilio, J., Pinto, P. A., \& Eastman, R. G. 1994, MNRAS, 266, L17
\item Suntzeff, N. B. 1996, IAU Colloquium 145: Supernovae and Supernova Remnants, ed. R. McCray, (Cambridge: Cambridge University Press), 41
\item Tammann, G. A. 1978, Astronomical Uses of the Space Telescope, eds. F. Macchetto, F. Pacini \& M. Tarenghi (Garching: ESO Proceedings), 329
\item Tammann, G. A., this volume
\item Teerikorpi, P. 1998, A\&A, 339, 647
\item Turatto, M., Benetti, S., Cappellaro, E., Danziger, I. J., Della Valle, M., Gouiffes, C., Mazzali, P. A., \& Patat, F. 1996, MNRAS, 283, 1
\item Wambsganss, J., Cen, R., \& Ostriker, J. P. 1998, ApJ, 494, 29
\item Wambsganss, J., Cen, R., Xu, G., \& Ostriker, J. P. 1997, ApJ, 475, L81
\item White, M. 1998, ApJ, 506, 495
\item Wilson, O. C. 1939, ApJ, 90, 634
\item Woosley, S. E., \& Weaver, T. A. 1994, ApJ, 423, 371
\item Woudt, P., Leibundgut, B., \& Spyromilio, J. 1999, Wide-Field Surveys in Cosmology, eds. Y. Mellier and S. Colombi, in press

\end{list}

\end{document}